\documentclass[aps,pre,twocolumn,eqsecnum]{revtex4}

\usepackage{amsmath,bm,epsfig,here}

  \def\Fbox#1{\vskip1ex\hbox to 8.5cm{\hfil\fboxsep0.3cm\fbox{%
    \parbox{8.0cm}{#1}}\hfil}\vskip1ex\noindent}  %%  {TEXT} in BOX

\newcommand{\B}[1]{{\bm{#1}}}%% Bold Roman & Greek Lower & Upper Case
\newcommand{\C}[1]{{\mathcal{#1}}}    %%   Calligrapfic Upper case
\newcommand{\BC}[1]{\bm{\mathcal{#1}}}%% Bold Calligrapfic Upper case
\renewcommand{\sb}[1]{_{\rm{#1}}}  %% sub-   for lower case
\newcommand{\Sb}[1]{_{\scriptscriptstyle\rm{#1}}} %% Sub-   for Upper case
\newcommand{\eq}[1]{(\ref{#1})}%%  requires \eq{label}
\newcommand{\Eq}[1]{Eq.~(\ref{#1})}%%  requires \eq{label}
\newcommand{\Eqs}[1]{Eqs.~(\ref{#1})}%%  requires \eq{label}
\begin{document}

\title{Equations of motion and conservation laws in a theory of  stably stratified turbulence }

\author{Victor S. L'vov and Oleksii Rudenko}
\email{oleksii@wisemail.weizmann.ac.il}%%
\affiliation{Dept. of Chemical Physics, The Weizmann Institute of
Science, Rehovot 76100, Israel}

\begin{abstract}
  The letter considers non-isothermal fluid flows and revises simplifications of basic hydrodynamic equations for such flows arriving eventually to a generalization of the Oberbeck-Boussinesq approximation valid for arbitrary equation of state including both non-ideal gases as well as liquids. The proposed approach is based on a suggested general definition of  potential temperature. Special attention is put on the energy conservation principle, and it is shown that the proposed approximation exactly preserves the total mechanical energy by approximate equations of motion. The principal importance for any turbulent boundary layer model to respect the conservation laws is emphasized explicitly.
\end{abstract}

 % \keywords{Atmospheric Boundary Layer, Richardson Number, Transport Equations, Stratification}

\maketitle

\noindent {\textbf{Introduction}} \\ \vskip -.5cm

\noindent  In this letter, which has partially a pedagogical character,  we reconsider  simplifications of the basic hydrodynamic equations of motion for non-isothermal fluid flows, which can serve as a basis for a consistent theory of atmospheric turbulent boundary layers with stable temperature stratification. The reason for doing so stems from the fact that textbooks derivations of the celebrated Oberbeck-Boussinesq approximation  are too specialized either to
fluids which density variation is very small or to ideal gases and cannot be applied,  for example,  to humid air. Our generalization of the Oberbeck-Boussinesq approximation is based on a suggested below generalized definition of the potential temperature, which is valid for arbitrary equation of state, while  its standard  definition, widely used in meteorology, is applicable only for ideal gases.

We  discuss also a form of mechanical energy (consisting of the kinetic and potential energies) that is exactly preserved by the resulting approximate equations of motion. This integral of motion plays a crucial role in modeling  stably stratified  atmospheric turbulent boundary layers (see our paper in the same issue), allowing to control further approximations with the goal of a reasonable, simple, but still adequate description of turbulent fluxes that determine basic physics of the turbulent boundary layers.   \\

%%%%%%%%%%%%%%%%%%%%%%%%%%%%%%%%%%%%%%%%%%%%%
\noindent  {\textbf{1. Generalization of the Oberbeck-Boussinesq approximation for non-ideal gases and liquids}} \\

\noindent  {\textbf{1A.   Basic hydrodynamic equations}}.
The system of hydrodynamic equations describing a fluid in which the temperature
is not uniform  consists of the
Navier-Stokes equation for the fluid velocity, $\BC U(\B r,t)$, a continuity
equation for the space and time dependent (total) density of the fluid, $\rho
(\B r,t)$, and of the heat balance equation for the (total) entropy
per unit mass, $\C S(\B r,t)$,~\cite{LL}: %%  Landau and Lifshitz, 1987
\begin{subequations}\label{full}%%
\begin{eqnarray} %%
 \rho{\left({\!\frac{\partial }{\partial t} + \BC U \cdot \B {\nabla}\!}\right)}\BC U &=&  - \B {\nabla}
 p  + \B g\, \rho  +  \B \nabla \cdot \mu\, \B \nabla\BC U\;,~~~~~~~~~~
\label{B1} \\ %%
  \frac{\partial{\rho}}{\partial{t}}&\!\!+\!\!&
\B \nabla\cdot  \left(  \rho \, \BC U \right)   =  0\,,\label{B2}\\
%\end{eqnarray}
%\begin{eqnarray} %%
\label{LD}
  \rho \Big( \frac{\partial }{\partial t} &+& \BC U \cdot \B {\nabla} \Big)\C S = \B \nabla  \cdot \kappa\, \B \nabla \C S   \ .
 \label{B3}
\end{eqnarray}
\end{subequations}%%
 Here $p$ is the pressure, $\B g=-\widehat{\bf{z}}  g$ is the vertical
acceleration due to gravity, $\mu$ and $\kappa$ are the (molecular)
dynamical viscosity and heat conductivity.

These equations are considered with boundary conditions that
maintain the solution far from the equilibrium state, at which $\BC
U=\C S=0$. These boundary conditions are $\BC U=0$ at zero
elevation, $\BC U=const$ at a high elevation of a few kilometers.
This reflects the existence of a wind at high elevation, but we do
not attempt to model the physical origin of this wind in any detail.
The only important condition with regards to this wind is that it
maintains a momentum flux towards the ground that is prescribed as a
function of the elevation. Similarly, we assume that a stable
temperature stratification is maintained such that the heat flux
towards the ground is prescribed as well. In the entropy  balance
\Eq{B3} we have already neglected the viscous entropy production
term, $\propto \mu \B |\nabla \BC U|^2$, assuming that the
temperature gradients are large enough such that the thermal entropy
production term on the RHS of \Eq{B3} dominates. Actually, this
assumption is very realistic in meteorological and oceanographic applications. For simplicity of
the presentation we restrict ourselves by relatively small
elevations and   disregard in \Eq{B1} the Coriolis force (for more
details, see \cite{Wyngaard}).

On the other hand we assume that the temperature and density
gradients  in the entire turbulent boundary layer are sufficiently
small to allow employment of local thermodynamic equilibrium. In
other words, we assume the validity of the equation of state, and
that the entropy $\C S$ is a state function of the local values of
the density and pressure:
 \begin{equation}\label{B4}%%
 \rho =\rho(T , p)\,,\quad
\C S=\C S (\rho, p)\ . %%
\end{equation}%%
In the same manner we will neglect the temperature
dependence of the dissipation parameters $\nu$ and $\kappa$.

Pressure fluctuations caused by  turbulent velocity fluctuations $\B
u$ propagate in a compressible medium with the sound velocity $c\sb
s$,  causing time dependent density fluctuations of the order of
$(u/c\sb s)^2 \rho_0$, where $\rho_0$ is the mean density. Assuming
that the square of the turbulent Mach number $M\Sb T^2 \equiv  (u/c\sb
s)^2$ is small compared to unity, we can neglect in \Eq{B2} the
partial time derivative, hence %%
 $ \B \nabla\cdot  \left(  \rho \, \BC U \right) =0$ (see e.g. \cite{LL}).
  Even tropical hurricanes of category five have the mean wind
velocity $U$ is below 300 Km/h. Usually, the turbulent velocity
fluctuations $u$ are less then $U/10$, i.e. even in these extreme
conditions $u< 30$ Km/h and $M\Sb T^2< 10^{-3}$ (with $c\sb
s\simeq 1200$ Km/h). Therefore the incompressibility approximation $
\B \nabla \cdot \left(  \rho \, \BC U \right) =0$  is  well justified in atmospheric physics. In the ocean where
the sound velocity is even larger and water velocities even smaller,  this
approximation  is quite excellent.\\
%%%%%%%%%%%%%%%%%%%%%%%%%%%%%%%%%%%

\noindent{\textbf{1B. Isentropic basic reference state}}.
In quite air, without turbulence, the pressure and the density
depend on the elevation $z$ simply due to gravity. For example, in
full thermodynamic equilibrium the temperature is uniform,
$z$-independent, and the density decreases exponentially with the
elevation. However, this equilibrium  model of the atmosphere is not
realistic, and cannot be used as a reference state about which the
actual dynamics is considered. A much better reference suggestion is a
state in which the entropy is space homogeneous. In this model
the thermal conductivity (leading to the temperature homogeneity) is
neglected with respect to heat transfer due to the vertical
adiabatic mixing of air, leading to a $z$-independent entropy.
We refer to the isentropic  model as a ``basic
reference state" and denote this state of the system with a
subscript `` $\sb b$": %%
\begin{equation}\label{A3}
 \C S\sb
b=\C  S(\rho\sb{\, b}, p\sb {\, b} )={\rm const}\,, \quad
 \rho\sb{\, b} =\rho(T\sb b, p\sb {\, b} )\ . %%
\end{equation} %%
The first of \Eq{A3} relates the gradients of the pressure and
density in this state:%%
\begin{subequations}\label{rel1}\begin{equation}\label{rel1a} 0=\B \nabla \C S\sb b
=  \left( \frac{\partial  \C S\sb b}{\partial  \rho\sb b} \right) _p \B \nabla \rho\sb b+
 \left( \frac{\partial  \C S\sb b}{\partial  p\sb b} \right) _\rho  \B \nabla p\sb b \ .
\end{equation} Another relation between $\rho\sb b$ and $p\sb b$
follows from the condition of hydrostatic equilibrium:%%
 \begin{equation}\label{heq} %%
 \B \nabla  p\sb b=\B g \rho\sb b\ . %%
 \end{equation} \end{subequations} %%
 Equations \eq{rel1} together with the first of Eqs.~\eq{A3} determine the
density, pressure and temperature profiles in the isentropic basic reference state.\\

%%%%%%%%%%%%%%%%%%%%%%%%%
 \noindent{\textbf{1C. Hydrodynamic equations in generalized
 Oberbeck--Boussinesq approximation}}\\
\paragraph{Equations of motion.}
Denote the deviations of the total density, pressure, temperature and entropy
from the basic reference state as follows:%%
 \begin{eqnarray}\label{dev}  \hat\rho \equiv  \rho-\rho\sb{\,
b}\,, && \hat p\equiv p-p\sb{\, b}\,,  \\ \nonumber \hat T \equiv  T-T\sb{\, b}\,, &&
\hat{\C S}\equiv \C S-\C S\sb{\, b}\,. \end{eqnarray} Following
Oberbeck \cite{Oberbeck} and Boussinesq \cite{Boussinesq}, assume that these
deviations are small: %%
 $
\hat  \rho \ll \rho\sb{\, b}\,, \quad
\hat  p \ll p\sb{\, b}$.
Then  one   simplifies the full system of
 hydrodynamic Eqs.~\eq{full} and  rewrites them in terms of the
 fluid velocity $\BC U$ and the small deviations   $\hat p$ and $\hat {\C S}$
 (instead of $\hat \rho$). The first step is very simple:
 because of \Eq{heq}%%
 \begin{subequations}\label{step1}\begin{equation}\label{step1a}%%
  \B \nabla  p-  \B g \rho=
   \B \nabla \hat p -  \B g\, \hat \rho \ .
 \end{equation}
Next we should relate the deviations  $\hat p$, $\hat \rho$ and
$\hat {\C  S}$. In the linear approximation \Eq{B4} yields:%%
\begin{equation}\label{step1b} \hat{\C S}= \left(  \partial  \C S \sb b\big / \partial  p\sb {\, b} \right) _\rho
\hat p+  \left(  \partial  \C S \sb b\big / \partial  \rho \sb {\, b}  \right) _p \hat \rho\ .
\end{equation} %%
With the help of Eqs.~\eq{rel1} this gives: %%
\begin{equation}\label{step1c} %%
\B g \, \hat \rho =  \frac {\B \nabla \rho\sb b } {\rho\sb b}\,\hat
p  - \B  \beta\,  \frac{\rho\sb   b }{c\sb p}\, T\sb b \, \hat{\C
S}\ . %%
\end{equation}%%
Here $\B \beta\equiv  \B g \, \widetilde  \beta$ is the buoyancy parameter, $\widetilde \beta$
is the thermal expansion coefficient and $c\sb p$ is the isobaric
specific heat in
the basic reference state (for thermodynamic relations see, for example, \cite{Webster}):%%
\begin{equation} \label{beta} %%
\widetilde \beta  \equiv   - \frac{1}{\rho\sb b}  \left( \frac{\partial  \rho\sb b}{\partial  T\sb
b} \right) _p= -\frac {c\sb p }{\rho\sb b\, T\sb b}  \left( \frac{\partial  \rho \sb b}
 {\partial  \C S\sb b} \right) _p\ .
 \end{equation} %%
 Now \Eq{step1a} in the linear approximation yields:
\begin{equation}\label{step1e}%%
  \B \nabla  p-  \B g \rho=
   \rho\sb b\left[\B \nabla  \left(  \frac {p\sb b}{\rho\sb b } \right)  + \B \beta \,
   \frac{  T\sb b}{c_p} \, \hat{\C S}
  \right ]\ .
 \end{equation}
\end{subequations}%%
 Then Eqs.~\eq{B1} can be approximated as: %%
 \label{B6} %%
 \begin{equation}\label{B6a}%%
   \left({\!\frac{\partial }{\partial t}\! + \BC U\! \cdot\! \B {\nabla}\!}\right)\!\BC U   = -  \B {\nabla}\!\!  \left( \frac{ \hat p}{ \rho\sb b} \right)\!  - \B \beta \frac{  T\sb b}{c_p} \, \hat {\C S}  +\! \frac{1}{\rho_b}\B \nabla\! \cdot \mu\B \nabla \BC U\,.
 \end{equation} %%
%%%%%%%%%%%%%%%%%%%%%
\paragraph{Generalized potential temperature.}

To proceed, we generalize the notion of potential temperature
$\Theta$ (see, e.g. \cite{Stull}) which is traditionally defined as the temperature that a
volume of \emph{dry air} at a pressure $p(z)$ and temperature $T(z)$
would attain when adiabatically compressed to the pressure $p_*$
that exists at zero elevation $z=0$. This potential temperature can
be explicitly computed for an ideal gas with the result
\begin{equation}%%
\label{Thetaa} %%
  \overline{\Theta}(z)\equiv  T_* \big(p_*/p(z)\big) ^{(\gamma-1)/\gamma}\ ,%%
\end{equation} %%
where $\gamma$ is the ratio of
isobaric to isochoric specific heats, $\gamma\equiv c\sb p/c\sb v$, and
$T_*$ is the temperature at zero elevation.

We want to generalize the notion of the potential temperature for an
arbitrary stratified fluid requiring that in the isenotropic basic
reference state it would be constant $\Theta_*=T_*$. A second
requirement is that the definition will agree with Eq. (\ref{Thetaa})
for an ideal gas. Accordingly we define %%
\begin{equation}%%
  \overline{\Theta}(z)= T_* \exp\left[ \left( \C S(z)-\C S_{\rm b}  \right) /c\sb p\right]\ .  %%
  \label{defTheta}
\end{equation}%%
For more details see also \cite{PotTemp}. Indeed, if we
employ the equation of state and the equation for the entropy of an
ideal gas, i.e.
 \begin{equation}\label{ideal}%%
 p = \rho\, T \,,\quad
\C S =    \ln \left({p^{c\sb v }}\big/{\rho^{c\sb p} }\right)+{\rm const} \ ,%%
\end{equation} %%
one can easily check that Eq. (\ref{Thetaa}) is recaptured.

\paragraph{Resulting equations.} For small  deviations of   $\Theta$ from the basic reference state
value $T_*$, i.e. up to the linear order, \Eq{defTheta}  gives:%%
\begin{equation} \label{ThetaD} %%
\Theta\sb{\, d}\equiv   \overline{\Theta} -T_* =  T_*\,
 \hat {\C S}\big / c\sb p \ . %%
 \end{equation}
Now we can present Eqs.~\eq{B1}, \eq{B2} (with $\partial  \rho \sb{\, t}/\partial
t=0$, as explained) and \eq{B3} as follows: %%
\begin{subequations}\label{FS} %%
\begin{eqnarray}\label{FSa} %%
 \left({\!\frac{\partial }{\partial t} + \BC U\! \cdot\! \B {\nabla}\!}\right)\!\BC U\! & =&\!   %%
-  \B {\nabla}\!\!  \left( \frac{p}{ \rho\sb b} \right)\!  - \B \beta  \,  \Theta\sb {d}
 + \nu\, \Delta\, \BC U \,,~~~~~~~~~~~  \\
\label{FSb}  %%
  && \hskip -0.15 cm \B \nabla \cdot   \big( \rho\sb b\,  \BC U  \big)   =  0\,, \\
\label{FSc}%%
\Big(\frac{\partial }{\partial t}\! &+&\! \BC U \cdot \B {\nabla}\Big)\, \Theta\sb{d}  = \chi\, \Delta\,
\Theta\sb {d} \ .
\end{eqnarray}\end{subequations}%%
The dissipative terms are important only in the narrow region of the
viscous sublayer, where we can safely neglect the $z$-dependence of
$\rho \sb b$, $\mu$ and $\kappa$, and   consider the dynamical
viscosity  $\nu = \mu / \rho_b$ and dynamical thermal conductivity
$\chi = \kappa / \rho_b$ as some $z$-independent constants.

Note that for turbulence in liquids (water, etc.) one can simplify these equations
further. There one can neglect the effect of adiabatic cooling (together with
the compressibility), and simply use another reference state with constant
temperature and density:
\begin{equation}\label{BRS2} %%
T= T_*\,, \quad \rho\sb b= \rho_*\,, \quad p\sb b=p_*+ g \rho_* z\ .
\end{equation}%%
 For this reference state, the standard reasoning (see,
e.g. \cite{LL}) yields the same equations as \Eqs{FS} in which again
$\widetilde \beta$, is given by \Eq{beta} and it is a parameter characterizing
a particular fluid. In this case   $\rho\sb b=\rho_*$, independent
of $z$,  and the potential temperature $\Theta=T$, such that
$\Theta\sb {\, d}= T\sb {\, d}$ is a deviation of the total
temperature $T$ from its ground (bottom, or whatever) level $T_*$.

In the suggested \Eq{FS} the situation is more general since  we do not assume
that the reference state has the simple form~\eq{BRS2} (with $\rho\sb
b=\,$const). Importantly, on the RHS of \Eq{FSa} the density
$\rho\sb b(z)$ is operated on by the gradient, and the buoyancy term
$-\B \beta \Theta\sb {\, d}$ involves $\Theta\sb {\, d}\ne  T\sb {\,
d}$, the deviation of the \emph{potential temperature} defined by
\Eq{ThetaD}. This definition for liquids has nothing in common with
the standard meteorological definition~\eq{Thetaa}. Notice also that
for an ideal gas  $\widetilde \beta=1/T$, and \Eq{FSa} coincides with that
suggested  in the book \cite{Kurbatsky}. \\

%%%%%%%%%%%%%%%%%%%%%%%%%%%%%%%%%%%%%%%%%%%%%
\noindent  {\textbf{2. Conservation of total mechanical energy}}\\

\noindent  {\textbf{2A. Dynamical integral of motion}}.
It is important to  realize that the \emph{approximate} Eqs. \eq{FS}
\emph{exactly} conserve  an \emph{approximate} expression for the
total mechanic energy of the system in the dissipation-less limit.
This total energy is the sum of the kinetic, $\C E\Sb K$, and the potential energy
$\C E\Sb P$ (calculated in the basic reference state):%
\begin{equation} \label{TotMechEn}%%
 \C E\Sb K\equiv \int d \B r\,\rho_b\,   \frac{|\BC
U|^2}{2}\,, \quad   \C E\Sb P\equiv \int d \B r \rho_b~\B \beta\cdot \B r~ \Theta_d \  .%%
\end{equation}%%
 One can check by direct substitution that this sum of energies is conserved by equations of motion
 (\Eqs{FS}) when $\nu=\chi=0$, i.e the sum $\C E\Sb K + \C E\Sb P$ is a dynamical integral of motion.\\
 %%%%%%%%%%%%%%

\noindent  {\textbf{2B. Statistical integral of motion}}.
  We show now that the potential energy of a stratified turbulent
  flow $\C E\Sb P$,  \Eq{TotMechEn}, can be presented in a turbulent regime  as a sum of
  (time-independent) potential energy of the basic reference
  state, $\overline{\C E}\Sb P$ and a ``turbulent" potential energy,
  associated with temperature fluctuations, $\widetilde {\C E}\Sb P$, as it was first mentioned by L.~F.~Richardson \cite{Richardson}.
  Actually, it is more instructive to discuss this issue in a more general case, when
  the stratification is caused by some ``internal" parameter of the
  fluid,  $\xi$, not necessarily the potential temperature. It can be
  salinity of water in a sea, humidity of air, a
  concentration of particles co-moving with the fluid as Lagrangian
  tracers, etc.

In the general case  then the equation for the potential energy of a
stratified fluid has the form:%%
 \begin{subequations} \label{def-PE}%%
 \begin{equation}\label{def-PEa} \C E\Sb P= g\int \rho  (\B
r)   z \ dx \, dy\, dz\,,%%
 \end{equation}
In the basic reference state, the  potential energy reaches its
minimum value referred to as  the basic  potential energy,
$\overline{\C E} \Sb P $:
\begin{equation}\label{BPE}%%
\overline{\C E} \Sb P = g\int \rho  \Sb b (z)   z \ dx \, dy\, dz\ .
\end{equation}%%
Clearly,
in the equilibrium the fluid density $\rho  \Sb b (z)$ decreases with the
elevation: $d\rho \Sb b / dz<0$.
In a turbulent state the density deviates from its reference value:  $\rho (\B
r,t)= \rho  \Sb b (z)+ \widetilde  \rho  (\B r,t)$ and the mean  potential energy $ \left\langle  \C
E\Sb P \right\rangle $ exceeds $\overline{\C E} \Sb P $:
\begin{equation}\label{apeb}
   \left\langle  \C E\Sb P  \right\rangle   = \overline{\C E} \Sb P + \widetilde {\C E} \Sb P \ .
\end{equation}\end{subequations}%%
We compute the turbulent potential energy, $\widetilde {\C E}\Sb P $, in
the case when the internal parameter $\xi$ (temperature, etc.) is
co-moving with the fluid element as a Lagrangian marker.  In the
Lagrangian approach   we can consider $\rho $ as the Lagrangian marker
and introduce  a variable $z(\rho ,t )$, which is understood as an
elevation of the fluid element with the density $\rho $.  Noticing that
$zdz=\frac 12 dz^2$, and integrating  \Eq{def-PEa}  by parts with
respect of $z^2$,  we can present $ \left\langle \C E\Sb P \right\rangle $   in the
Lagrangian approach as:%%
 \begin{equation}\label{PE}  \left\langle \C E\Sb P \right\rangle  = -
\frac g2\int  \left\langle [z(\rho ,t )]^2  \right\rangle    dx \, dy\, d  \rho  \ . \end{equation}
 As a result of turbulent motion, the elevation $z(\rho ,t )$ at given $\rho $ fluctuates and  can be decomposed into the mean and fluctuating parts:%%
\begin{equation}%%
  \label{dec1} %%
  z(\rho ,t )=z \Sb b (\rho )+ \widetilde  z(\rho ,t )\,, \quad \left\langle \widetilde  z(\rho ,t) \right\rangle =0\ . %%
\end{equation} %%
The substitution of $z^2(\rho ,t )=z^2 \Sb b +2\, z \Sb b \widetilde z + \widetilde z^{\; 2}$ in \Eq{PE} leads to  three contributions to  the potential energy. The first one (originating from $z^2 \Sb b $) describes the basic
 potential energy, \Eq{BPE}. The second contribution, which is  linear in $\widetilde  z$, disappears
because $ \left\langle \widetilde z \right\rangle  =0$. The last one describes  the turbulent potential
energy:%%
\begin{equation}%%
  \label{PE1}
   \widetilde {\C E}\Sb P = - \frac g2\int  \left\langle [\widetilde  z (\rho ,t)]^2 \right\rangle     \, dx \, dy\, d\rho  \ . %%
\end{equation}
Relating  the density fluctuations, $\widetilde \rho $,  around the basic reference state density
profile $\rho  \Sb b (z)$ with $\widetilde  z$
 \begin{equation}\label{def-d}
\delta  \rho  \equiv  \frac{d \rho  \Sb b (z)}{d\, z}\, \delta  z\,, \end{equation} and
returning back to the Eulerian description in \Eq{PE1},
  one has:
\begin{equation}\label{tr1b}  \widetilde {\C E}\Sb P =  - \frac g2\int    \left\langle  \widetilde \rho  ^{\,2}
 \right\rangle  \Big[\frac{d\, \rho  \Sb b (z) }{d z  }\Big]^{-1}   dx \, dy\, dz \ .
\end{equation} Here we used the transformation formula, similar to \Eq{def-d}:
$d \rho  =  [d \rho\sb{b}(z)\big / d\, z ]\,  d z$. Equation~\eq{tr1b}
allows one to introduce a local density of  turbulent potential
energy per unit mass,
\begin{subequations}\label{def1}
\begin{equation}\label{def1a}
 E\Sb P  =  - \frac g2  \,  \Big[\frac{d\, \rho  \Sb b (z) }{d z  }\Big]^{-1} { \left\langle  {\widetilde \rho \,}^{2} \right\rangle }\Big/{\rho\Sb b}  \,,
\end{equation}
 such that
 \begin{equation}\label{def1b}\widetilde { \C E}\Sb P  =  \int
\rho\Sb b E\Sb P\, dx \, dy\, dz \ .
 \end{equation} \end{subequations}

For the particular case of temperature stratification %%
\begin{equation} \label{case}%%
\frac{d\, \rho  \Sb b
 (z) }{d z  } = \rho\Sb b \widetilde \beta \, \frac{d \Theta}{d z} \,,\quad \widetilde \rho =\rho\Sb b \widetilde \beta  \theta\,, \quad \beta =g \widetilde \beta \,,
 \end{equation}
where $\Theta = \left\langle\Theta\sb d\right\rangle$ is the mean potential temperature, $\theta  = \Theta\sb{\,d} -\Theta$ is the fluctuation of potential temperature, and the turbulent potential energy per unit mass for the   stratified turbulent boundary layers is:%%
\begin{equation} %%
\label{Ep-ddr}%%
  E\Sb P={\beta} E_\theta/{S_\Theta}\,,
\end{equation}
where $E_\theta = \left\langle \theta^2 \right\rangle\!/2$ and $S\Sb\Theta = d\Theta /d z$.

Clearly, a consistent statistical description of turbulent flows must conserve (in the dissipationless limit) the total mechanical energy, which for stratified flows consists of kinetic energy of the mean flow,  kinetic energy   of turbulent velocity fluctuations, $\bm u$, and turbulent potential energy, $E\Sb P$. To respect this conservation law one has to discuss explicitly  the balance equation for the turbulent kinetic energy that includes components of the Reynolds stress tensor $\langle u_i u_j \rangle$, and the balance equation for the potential energy, which is proportional to $\langle \theta ^2 \rangle$. Unavoidably, the turbulent heat flux vector $\langle \bm u \theta \rangle$ is involved into the game and requires a separate balance equation for itself. Therefore, a consistent statistical description of the stratified turbulent flows demands an explicit consideration of the whole set of the second-order one-point, simultaneous (cross)-correlation functions: $\langle u_i u_j \rangle$, $\langle \bm u \theta \rangle $ and   $ \langle \theta ^2 \rangle$.  Our version of such a consistent treatment is presented elsewhere in this issue.

\noindent  {\textbf{Summary}}. We presented here a generalization of the Oberbek-Boussinesq approximation for temperature stratified flows. We showed via a detailed discussion that this approach is acceptable for fluids where the local thermodynamic equilibrium is established i.e. the equation of state is valid. Thus the applicability of the approximation is much wider than for just ideal gases or liquids. For example, it is well established for humid air accounting for which is important for realistic meteorological prognoses, for salt water, hence oceanography, or even for fluids near the critical point met in engineering application. The paper also shows that the proposed generalization respects the conservation of the total mechanical energy. Obeying the conservation laws is of principle importance in a construction of consistent models, and a violation of this rule may lead to unphysical predictions. For stratified turbulent boundary layers this requirement lays in an explicit consideration of not only the mean profiles, but also of {all} relevant second-order, one-point, simultaneous correlation functions of  {all} fluctuating fields.\\
\noindent {\bf Acknowledgement}. We acknowledge the
support of the Transnational Access Programme at RISC-Linz, funded
by European Commission Framework 6 Programme for Integrated
Infrastructures Initiatives under the project SCIEnce (Contract No.
026133).

%%%%%%%%%%%%%%%%%%%%%%%%%%%%%%%%%%

%

\end{document}